\documentclass{elsart}

\usepackage{graphicx}

\usepackage{amssymb}

\begin{document}

\begin{frontmatter}



\title{Adaptation to synchronization in phase-oscillator networks}


\author{Fernando Arizmendi and Dami\'an H. Zanette\thanksref{label2}}
\thanks[label2]{Also at Consejo Nacional de Investigaciones
Cient\'{\i}ficas y T\'ecnicas, Argentina.}
\ead{arizmenf@ib.cnea.gov.ar} \ead{zanette@cab.cnea.gov.ar}
\address{Centro At\'omico Bariloche and Instituto Balseiro \\
8400 San Carlos de Bariloche, R\'{\i}o Negro, Argentina}

\begin{abstract}
We introduce an adaptation algorithm by which an ensemble of coupled
oscillators with attractive and repulsive interactions is induced to
adopt a prescribed synchronized state. While the performance of
adaptation is controlled by measuring a macroscopic quantity, which
characterizes the achieved degree of synchronization, adaptive
changes are introduced at the microscopic level of the interaction
network, by modifying the configuration of repulsive interactions.
This scheme emulates the distinct levels of selection and mutation
in biological evolution and learning.
\end{abstract}

\end{frontmatter}


\section{Introduction}

The emergence of coherent collective behaviour, out of the
interaction of a large number of active elements with relatively
simple individual dynamics, is the key signature of a complex system
\cite{csyst}. The nature of the mechanisms which underly such
emergence, however, can substantially vary between different
systems. In physical or chemical phenomena --for instance, in the
formation of spatiotemporal patterns in reacting fluids-- collective
behaviour is the unavoidable, spontaneous consequence of elementary
mechanical processes occurring at microscopic scales. In biological
systems, on the other hand, collective behaviour is not just the
macroscopic manifestation of the mutual organization of microscopic
entities. It is, as well, the consequence of a much longer
evolutionary process, by which certain macroscopic traits are
selected at the expense of others, while they change from one
generation to the next according the laws of heredity. The joint
action of natural selection and mutations, as described by the
Darwinian theory of evolution, makes it possible that, in a given
environment, the collective performance of a class of biological
complex systems --a species-- improves as generations succeed each
other. Similarly, the acquisition of certain forms of behaviour by
learning implies a process by which the response of a living being
to given stimuli becomes specific and optimized.

It is interesting that --both in evolution and in learning-- the
mechanism of selection, on one hand, and the changes that eventually
improve performance, on the other, act at very distinct levels.
Natural selection of beneficial traits, as well as evaluation of
learnt tasks, take place at macroscopic scales, at the level of the
organic interaction of individuals with their environment. Variation
processes, in contrast, take place at microscopic scales, genetic
for evolution and neural for learning. It is precisely the emergent
nature of macroscopic traits which connects the two levels.

In this paper, we explore the possibility of emulating this scenario
with a rather simple model, based on the collective dynamics of
coupled oscillators. It is well known that ensembles of interacting
oscillators may undergo self-organization processes which lead to
different forms of synchronized behaviour \cite{Kura,Manr}. A
variety of collective states --in the form of single or multiple
synchronized clusters, for instance-- can be realized by altering
the pattern of interactions between oscillators. We exploit this
behavioral diversity for inducing an oscillator ensemble to adopt a
specific, target synchronization state, through adaptive
introduction of gradual changes in its interaction pattern.
Imitating biological adaptation, achievement of the goal is
evaluated through a macroscopic measure of synchronization, while
changes are introduced at the microscopic level of pair
interactions.

We analyze, in the next section, the possible synchronization states
of an oscillator ensemble with both attractive and repulsive
interactions. We show that if repulsive interactions are
conveniently distributed, the ensemble splits into two synchronized
clusters with opposite phases. Choosing this two-cluster
configuration as the target state, in Section \ref{adapta} we apply
an adaptation algorithm to induce an ensemble, whose repulsive
interactions are initially distributed at random, to modify its
interaction pattern in such a way that its collective behaviour
approaches the target. This formulation contrasts with previous
studies of network adaptation to self-organization \cite{zhou},
where the degree of synchronization is evaluated at local --not
collective-- level. In spite of the huge number of possible ways of
distributing the repulsive interactions, our system manages to
satisfactorily achieve its goal.

\section{Oscillator networks with repulsive interactions}

We consider a Kuramoto-like model for $N$ coupled phase oscillators,
governed by the equations \cite{Kura,Daido1}
\begin{equation} \label{model}
\dot \phi_i = \omega_i + \frac{K}{N} \sum_{j=1}^N W_{ij}
\sin(\phi_j-\phi_i),
\end{equation}
where $\phi_i (t) \in [0,2\pi)$ is the phase of oscillator $i$, and
$\omega_i$ is its natural frequency. The positive constant $K$
measures the coupling between oscillators, and $W_{ij}=\pm 1$
defines the sign of the interaction between oscillator $i$ and $j$.
Positive and negative $W_{ij}$ represent, respectively, attractive
and repulsive ``forces'' between oscillators. Interactions are
assumed to be symmetric, so that $W_{ij}=W_{ji}$. In the following,
it will be useful to conceive the interaction pattern as defining a
network with vertices occupied by the oscillators, whose links join
oscillators with attractive interactions. The complementary network,
with links corresponding to repulsive interactions, provides an
alternative representation \cite{z1}.

It is well-known that in the case of global attractive coupling,
$W_{ij}=1$ for all $i$ and $j$, and in the thermodynamical limit, $N
\to \infty$, the oscillator ensemble undergoes a synchronization
transition as the coupling strength $K$ grows \cite{Kura}. This
transition is quantified by the Kuramoto order parameter
\begin{equation} \label{z}
z= \frac{1}{N} \left< \left| \sum_{j=1}^N \exp ({\rm i}\phi_j)
\right| \right>,
\end{equation}
where $\left< \cdot \right>$ indicates averages over sufficiently
long times. Below the critical coupling strength $K_c$, phases are
uniformly distributed in $[0,2\pi)$, so that $z=0$. Above $K_c$, on
the other hand, $z$ grows monotonically with $K$, revealing a
progressive condensation of phases. At the same time, an
increasingly large cluster of synchronized oscillators, whose
time-averaged frequencies,
\begin{equation} \label{Omega}
\Omega_i = \left< \dot \phi_i\right> = \lim_{T\to \infty}
\frac{1}{T} \int_0^T \dot \phi_i (t) dt,
\end{equation}
are mutually identical, dominates the collective dynamics of the
ensemble. The critical point $K_c$ depends on the distribution of
natural frequencies $\omega_i$: a higher dispersion in the natural
frequencies requires a stronger coupling to synchronize the
ensemble. For asymptotically large $K$, the cluster of synchronized
oscillators entrains the whole ensemble, and all the phases collapse
to the same value, so that $z \to 1$.

\begin{figure}[h]
\begin{center}
\resizebox{\columnwidth}{!}{\includegraphics{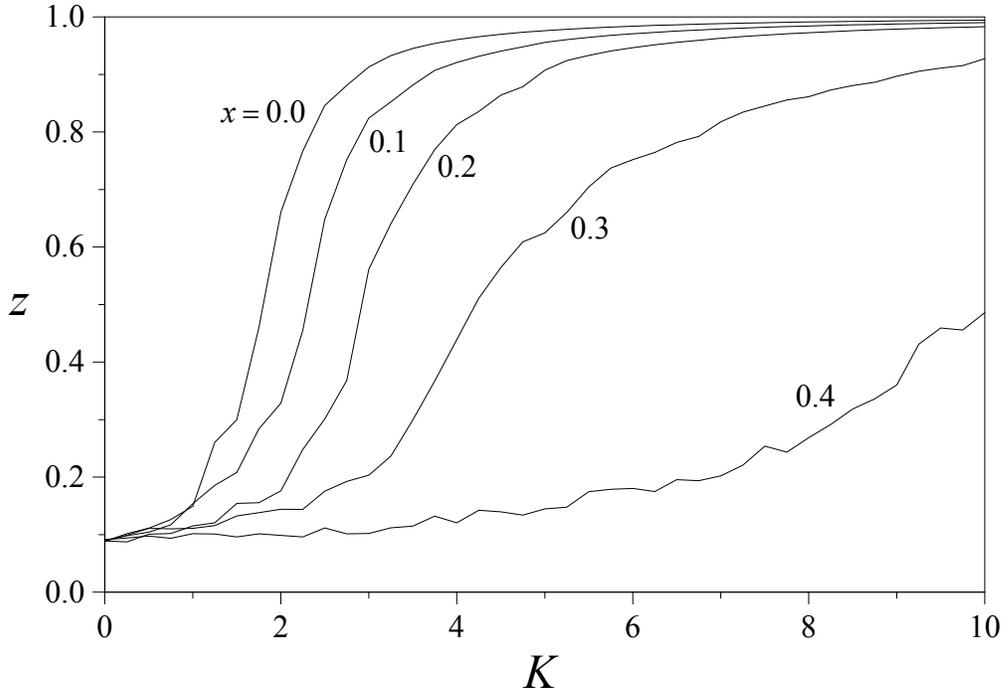}}
\end{center}
\caption{The Kuramoto order parameter $z$  as a function of the
coupling strength $K$, for an ensemble of $10^3$ oscillators with a
fraction $x$ of repulsive interactions distributed at random.
Repulsive interactions define an Erd\H{o}s-R\'enyi network over the
oscillator ensemble.} \label{f1}
\end{figure}

As it may be expected, when some of the interactions are repulsive,
synchronization is harder to attain. This was originally confirmed
by Daido \cite{Daido2}, who studied Eqs. (\ref{model}) in the case
where the symmetric weights $W_{ij}$ are drawn at random from a
Gaussian distribution centered at zero. Moreover, it turns out that
the emerging patterns of collective behavior depend crucially on the
form in which repulsive interactions are distributed over the
ensemble. In Ref. \cite{z1}, the effect of a random
(Erd\H{o}s-R\'enyi \cite{ER}) network of repulsive interactions was
studied for identical phase oscillators ($\omega_i\equiv \omega$ for
all $i$). It was shown that, as the number of repulsive interactions
grows,  fully synchronized ensembles undergo a sharp transition to
desynchronization. Figure \ref{f1}, on the other hand, shows results
for ensembles of $N=10^3$ {\em nonidentical} oscillators with an
Erd\H{o}s-R\'enyi network of repulsive interactions. Natural
frequencies are drawn at random from a Gaussian distribution
$g(\omega) = \exp(-\omega^2/2)/\sqrt{2\pi}$. Each interaction weight
is given by
\begin{equation}
W_{ij}  = W_{ji} =\left\{
\begin{array}{rl}
1 & \mbox{with probability $1-x$,} \\ -1 &\mbox{with probability
$x$,}
\end{array}
\right.
\end{equation}
so that $x$ is the average fraction of repulsive interactions. The
plot displays the order parameter $z$, Eq. (\ref{z}), as a function
of the coupling strength $K$ for several values of the fraction $x$.
Note that $x < 1/2$ in all cases, so that the number of attractive
interactions is always larger than that of repulsive interactions.

From Kuramoto's theory \cite{Kura}, it is known that in the absence
of repulsive interactions ($x=0$) the synchronization transition for
the present distribution of natural frequencies takes place at
$K_c\approx 1.6$. Our results show that, as $x$ grows, the
transition shifts to higher values of $K$, and the transition zone
becomes broader. Nevertheless, for sufficiently large coupling
intensities, the order parameter seems to approach its maximum
value, $z=1$, which corresponds to the state of full
synchronization. Thus, as coupling becomes stronger, the more
abundant attractive interactions overcome the effect of repulsive
interactions, and the whole ensemble aggregates into a single
cluster with well-defined phase.

Other modes of collective behaviour can be expected if, instead of
being random, the networks of attractive and repulsive interactions
are given some special structure. Specifically, consider that the
oscillator ensemble is divided into two groups, such that inside
each group all interactions are attractive, whereas they are
repulsive between oscillators of different groups. In this
situation, the network of attractive interactions is fully connected
inside each group, and the two groups are mutually disconnected. The
interaction weights are given by
\begin{equation}\label{2cl} W_{ij} =\left\{
\begin{array}{rl}
1, & \mbox{if $i$ and $j$ belong to the same group,} \\ -1, &
\mbox{otherwise.}
\end{array}
\right.
\end{equation}
If the groups have, respectively, $N_1$ and $N_2$ oscillators
($N_1+N_2=N$), the fraction of repulsive interactions is
$x=2N_1N_2/N(N-1)$, or equivalently,
\begin{equation} \label{N12}
N_{1,2} = \frac{N}{2} \left[ 1\pm \sqrt{1-2x(N-1)/N}\right] \approx
\frac{N}{2} \left( 1\pm \sqrt{1-2x}\right) ,
\end{equation}
where the approximation holds for large $N$. Note that these numbers
make sense if $x<N/2(N-1)\approx 1/2$.

\begin{figure}[h]
\begin{center}
\resizebox{\columnwidth}{!}{\includegraphics{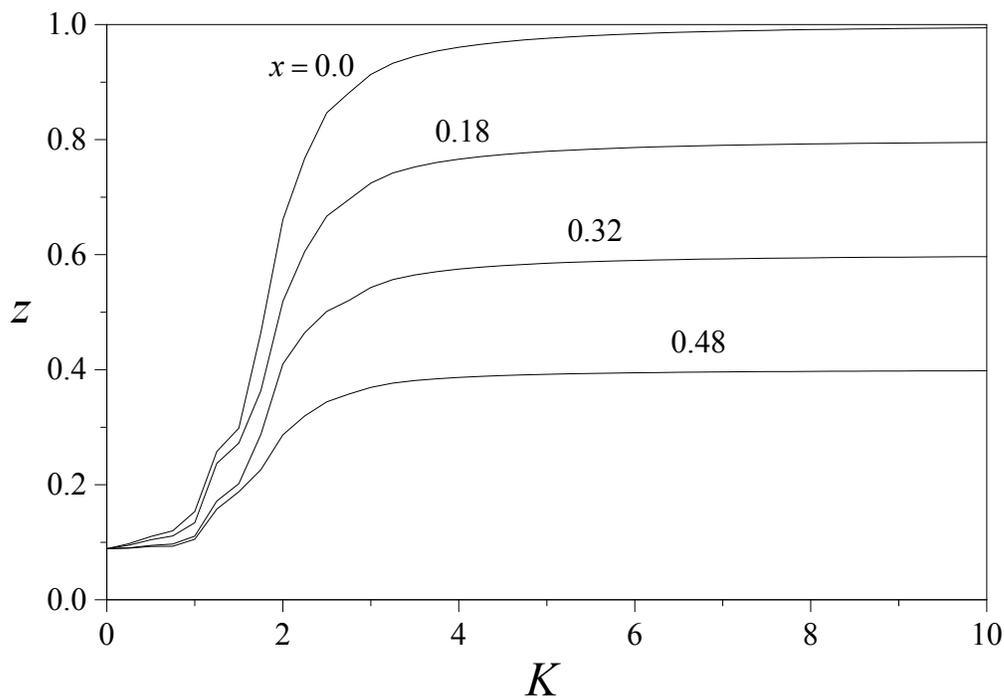}}
\end{center}
\caption{The Kuramoto order parameter $z$ as a function of the
coupling strength $K$, for an ensemble of $10^3$ oscillators divided
into two groups. The interaction between two oscillators is
attractive (repulsive) if they (do not) belong to the same group.
The sizes $N_1$ and $N_2$ of the two groups are given, as a function
of the fraction  $x$ of repulsive interactions, by Eq. (\ref{N12}).}
\label{f2}
\end{figure}

It can be easily realized that, if all natural frequencies are equal
($\omega_i\equiv \omega$ for all $i$), the only stable solution to
Eqs. (\ref{model}) with the  interactions given by Eq. (\ref{2cl})
corresponds to all the oscillators of each group having exactly the
same phase, and the two groups having opposite phases. In other
words, an ensemble of identical oscillators with the interactions of
Eq. (\ref{2cl}) splits into two point-like opposite clusters in the
phase coordinate, each cluster corresponding to one group. When
natural frequencies are not mutually identical, we expect that the
clusters spread in phase but, for large coupling strengths, the
division into two opposite phase clusters corresponding to the
groups should persist. This is in fact what numerical results show.
Figure \ref{f2} displays results for the order parameter $z$  as a
function of the coupling strength $K$ for several values of $x$,
with the distribution of interactions given by Eq. (\ref{2cl}), in a
system of $N=10^3$ oscillators. Note that, now, the  transition to
synchronized behaviour occurs always at the same value of $K$. On
the other hand, the asymptotic value of $z$ for large $K$ depends on
$x$. Indeed, as discussed above, for large $K$ the ensemble is
organized in two point-like clusters of opposite phase so that,
according to Eq. (\ref{z}), the order parameter equals
\begin{equation} \label{zinf}
z({K \to \infty})= \frac{|N_1-N_2|}{N}= \sqrt{1-2x(N-1)/N} \approx
\sqrt{1-2x} ,
\end{equation}
an analytical prediction which is in full agreement with numerical
results.

\begin{figure}[h]
\begin{center}
\resizebox{\columnwidth}{!}{\includegraphics{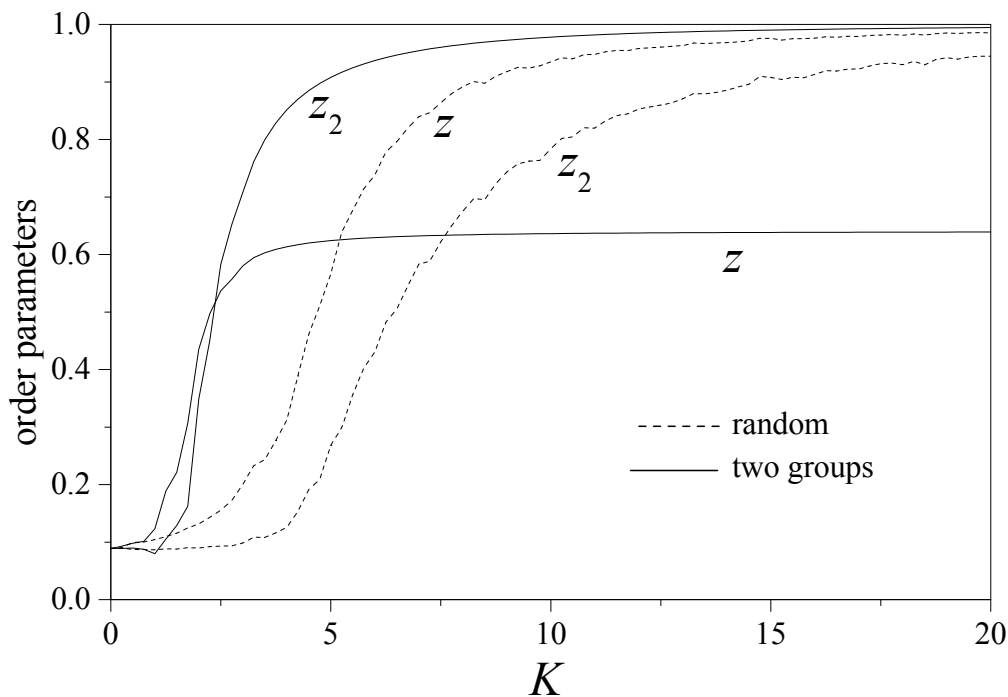}}
\end{center}
\caption{The order parameters $z$ and $z_2$ as functions of the
coupling strength $K$, for an ensemble of $50$ oscillators with a
fraction $x=0.3$ of repulsive interactions distributed at random
(dashed lines) or between two groups (full lines).} \label{f3}
\end{figure}

The splitting of the ensemble into two clusters is quantitatively
revealed by an additional order parameter \cite{Daido3},
\begin{equation} \label{z2}
z_2= \frac{1}{N} \left< \left| \sum_{j=1}^N \exp (2{\rm i}\phi_j)
\right| \right>
\end{equation}
[cf. Eq. (\ref{z})], which vanishes if oscillator phases are
uniformly distributed in $[0,2\pi)$, and attains its maximum,
$z_2=1$, if oscillators form two point-like clusters of opposite
phases. In this limit, $z_2$ is independent of the relative size of
the two clusters, so that $z_2=1$ also in the extreme case of a
single point cluster --equivalent to $N_1=N$ and $N_2=0$, or {\it
vice versa}. Therefore, for a random interaction network, both $z$
and $z_2$ approach unity as the coupling strength grows. It can be
easily shown that, in this case, $z>z_2$. For the two-group network,
on the other hand, $z_2$ tends to unity, while $z$ approaches the
asymptotic value quoted in Eq. (\ref{zinf}) so that, for large $K$,
$z<z_2$. This is illustrated in Fig. \ref{f3}, for a fixed value of
the fraction $x$ of repulsive interactions. The combined analysis of
$z$ and $z_2$ as a function of $K$ thus discerns between the two
different forms of collective behaviour of the oscillator ensemble,
derived from the different structures of their interaction networks.

\section{Adaptation to collective organization} \label{adapta}

In the preceding section, we have shown that a specifically
structured interaction network  induces splitting of the oscillator
ensemble into two synchronized clusters with opposite phases. In
contrast, a random distribution of attractive and repulsive
interactions gives rise to a single synchronized cluster. Suppose
now to have an ensemble with randomly distributed interactions.
Would it be possible, following our discussion in the Introduction,
to adaptively evolve the interaction pattern toward the two-group
structure by applying selection pressure at the level of collective
behaviour? This would amount to modifying microscopic attributes by
selection of macroscopic features, as in biological evolution and
learning.

In order to explore this possibility, we implement an algorithm of
adaptation of the interaction pattern based on a Monte Carlo scheme.
The algorithm is controlled by  monitoring the collective dynamical
state of the oscillator ensemble, as quantified by the two-cluster
order parameter $z_2$. We work at intermediate values of the
coupling strength, where the difference in $z_2$ between the two
dynamical modes studied above is larger (cf. Fig. \ref{f3}). In this
intermediate region, coupling is sufficiently strong to induce
synchronized collective behaviour, but does not reach the regime
where oscillators are confined to almost point-like clusters. In our
calculations, we choose $K=4$.

We start from an interaction pattern with a fraction $x$ of
repulsive interactions distributed at random all over the ensemble,
with weights $W_{ij}$. Equations (\ref{model}) are solved by means
of an Euler algorithm, with time increment $h=0.1$. The system is
left to evolve for  a few hundred time units (typically, $T=250$) so
that $z_2$ attains a well-defined value. Then, a small change is
introduced in the interaction pattern. Two oscillator pairs, $\{ k,l
\}$ and $\{ r,s \}$, with interactions of opposite signs, $W_{kl}=1$
and $W_{rs}=-1$, are chosen at random and their interaction weights
are interchanged, so that $W'_{kl}=-1$ and $W'_{rs}=1$. This
procedure is repeated on $n_P$ randomly chosen oscillator pairs.
With this new interaction pattern, the system evolves again and a
new value of the two-cluster order parameter, $z_2'$, is determined.
If $\Delta z_2=z'_2-z_2>0$, the change in the interaction weights is
accepted. If on the other hand, $\Delta z_2<0$,  the change is
accepted with probability $\exp(\Delta z_2 / \theta)$, and rejected
otherwise. The fictitious temperature $\theta$ is used to control
the speed of the adaptation process. The algorithm is repeated a
number of times (typically, several hundred), until the successive
values of $z_2$ converge to a stationary level. Following the
standard Monte Carlo practice, the temperature $\theta$ is slowly
decreased as the process progresses, to gradually attenuate the
effect of fluctuations in $z_2$. In our simulations, the number of
pairs whose interaction weights are modified at each adaptation step
is around $n_P=10$, and is adjusted to make the final value of $z_2$
as large as possible.

To illustrate our results, we consider a realization of the above
Monte Carlo process in an ensemble of $N=50$ oscillators, where the
fraction of repulsive interactions is $x=3/7\approx 0.43$. In a
two-group interaction network, this would correspond to groups of
sizes $N_1=15$ and $N_2=35$. From an initial value $z_2\approx
0.14$, obtained for a random interaction network, the two-cluster
order parameter significantly increases by a factor slightly above
four, reaching $z_2\approx 0.57$. Numerical results for systems of
this size, with $K=4$ and the two-group structure given by the
interaction weights of Eq. (\ref{2cl}), show however that the
two-cluster order parameter can reach values around $z_2=0.9$. After
the Monte Carlo process, the obtained value of $z_2$ is still
considerably below this level. In any case, in order to evaluate to
which extent has the system evolved towards the target state, we
examine both its resulting dynamical behaviour and its interaction
pattern.

\begin{figure}[h]
\begin{center}
\resizebox{\columnwidth}{!}{\includegraphics{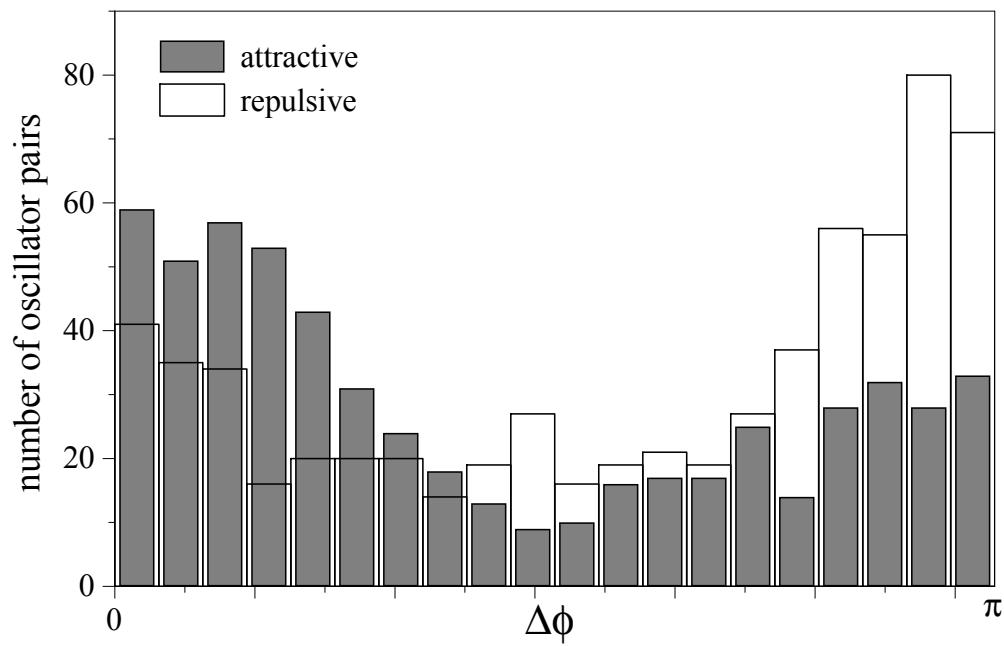}}
\end{center}
\caption{Number of oscillator pairs $\{ i, j \}$ with relative phase
$\Delta \phi_{ij}=|\phi_i-\phi_j |$, and attractive or repulsive
interactions. Individual phases were measured at a fixed time after
adaptation to the two-group state, in a realization with the
parameters quoted in the text.} \label{histo}
\end{figure}

\subsection{Distribution of phases and synchronization} \label{ss1}

Figure \ref{histo} shows results illustrating the correlation
between the relative phase of each oscillator pair and the sign of
the corresponding interaction, after adaptation. For two oscillators
$i$ and $j$, the relative phase $\Delta \phi_{ij}=|\phi_i-\phi_j |$
should be close to $0$ or $\pi$ for $W_{ij}=1$ or $-1$,
respectively. Indeed, as expected, the histograms of Fig.
\ref{histo} display a clear excess of attractive (repulsive)
interactions for small (large) relative phases. A quantitative
measure of the correlation between $\Delta \phi_{ij}$ and $W_{ij}$
is given by the index
\begin{equation} \label{C}
C = \frac{2}{N(N-1)} \sum_{i,j>i} W_{ij} \cos \Delta \phi_{ij}.
\end{equation}
It is easily shown that, for the synchronized state with two point
clusters --which would be achieved for large coupling strengths and
a perfect two-group interaction network-- we get $C=1$. When
oscillators are  uniformly spread in phases, on the other hand, $C
\sim N^{-1}$. The latter is the expected value of $C$ for a random
interaction network. In fact, from Fig. \ref{f1}, we note that for
the present values of $x$ and $K$ the oscillator ensemble with
randomly distributed repulsive interactions is essentially
unsynchronized. For the data corresponding to Fig. \ref{histo}, in
contrast, we find $C\approx 0.22$. As may have been expected, due to
the moderate coupling strength, this is still far from the maximum.
However, $C$ is an order of magnitude larger than the value expected
for an unsynchronized ensemble. This indicates that adaptation has
had a substantial effect in the desired direction.

\begin{figure}[h]
\begin{center}
\resizebox{\columnwidth}{!}{\includegraphics{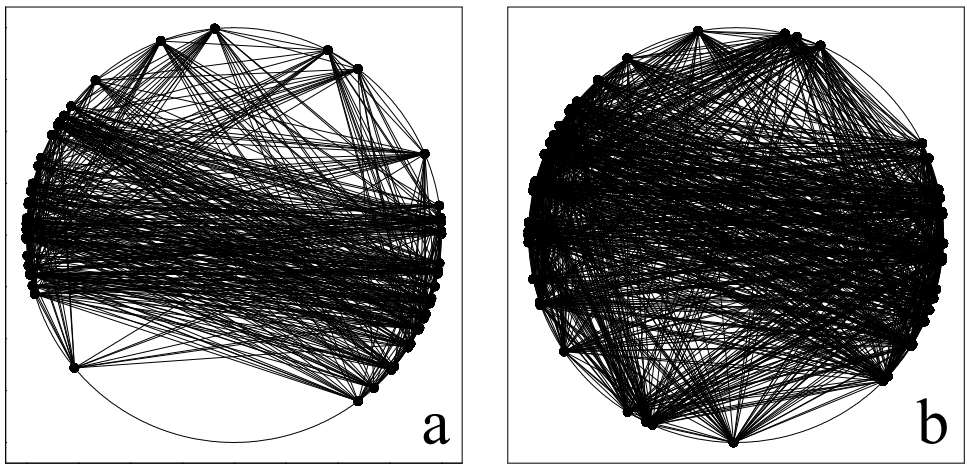}}
\end{center}
\caption{Two snapshots of the distribution of phases after
adaptation to the two-group state, in an ensemble of $50$
oscillators with the parameters specified in the main text.
Individual phases are represented over the unit circle, and lines
stand for repulsive interactions between oscillators.} \label{f4}
\end{figure}

Closer inspection of the distribution of phases reveals a
complementary consequence of the moderate value of $K$. The two
plots of Fig. \ref{f4} are representations of the individual phases
of the $50$ oscillators over the unit circle. Lines joining
oscillators represent repulsive interactions. Each plot corresponds
to a different time, i.e. it shows a snapshot of the distribution of
phases. In Fig. \ref{f4}a, it is clear that most of the ensemble has
split into two clusters with approximately opposite phases, with
most of the repulsive interactions standing between oscillators in
different clusters. Some time later, as illustrated by the snapshot
of Fig. \ref{f4}b, clusters are however much less defined and the
distribution over the unit circle is  more uniform. While most of
the repulsive interactions still correspond to pairs of oscillators
with opposite phases, it is clear that the ensemble is considerably
more spread.

\begin{figure}[h]
\begin{center}
\resizebox{\columnwidth}{!}{\includegraphics{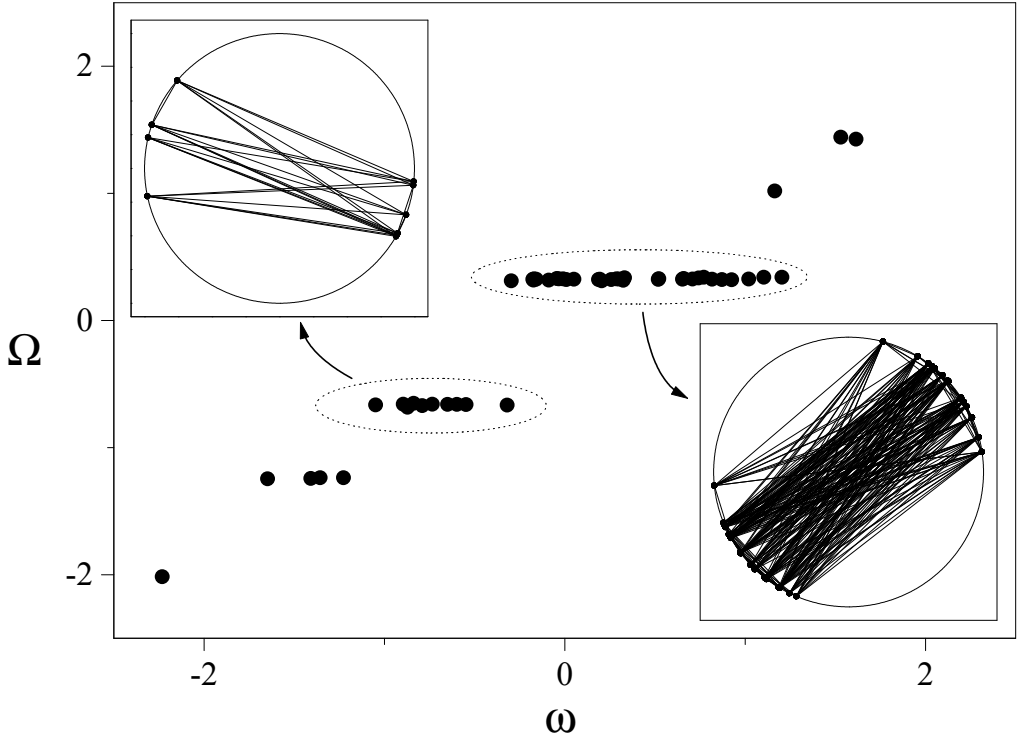}}
\end{center}
\caption{Main plot: The average frequency $\Omega_i$ as a function
the natural frequency $\omega_i$ for individual oscillators after
adaptation, with the parameters specified in the main text. Inserts:
Snapshots of the distribution of phases (cf. Fig. \ref{f4}) for the
synchronized clusters inside dotted frames.} \label{f5}
\end{figure}

To disclose the origin of this apparent time variation in the degree
of organization in phases, it is useful to consider as well the
distribution of frequencies over the ensemble. The main plot in Fig.
\ref{f5} represents the average frequency $\Omega_i$, defined in Eq.
(\ref{Omega}), as a function of the natural frequency $\omega_i$ for
each individual oscillator. In this kind of plot, horizontal arrays
of dots represent clusters of oscillators whose natural frequencies
are different, but which have synchronized to a common average
frequency. At the present value of $K$, we find several of these
clusters, of different sizes, with different average frequencies.
The analysis of the distribution of repulsive interactions inside
each synchronized cluster shows that the adaptation algorithm has in
fact succeeded at separating groups with opposite phases. This is
displayed for two of the synchronized clusters in the inserts of
Fig. \ref{f5}. However, since each synchronized cluster has its own
average frequency, they move around the unit circle with respect to
each other. Thus, their relative position changes, and the
clustering in the distribution of phases is at times more evident,
at times less defined. This observation brings to light a subtle
difference between the goal of the adaptation algorithm and the
synchronization of the ensemble: for intermediate values of the
coupling strength, pairs of clusters with opposite phases form due
to the redistribution of repulsive interactions, though
desynchronization between different pairs may persist.

\subsection{Structure of the interaction network}

The results discussed in Section \ref{ss1} allow us to appraise the
performance of adaptation at the level of the dynamical state
achieved by the ensemble, through the correlation between phases and
the distribution of repulsive interactions, and through the degree
of synchronization. This macroscopic viewpoint can be complemented
by an evaluation of the microscopic structure of the resulting
interaction network. In other words, we aim at evaluating to which
extent the redistribution of interactions has transformed the
initial random structure into the two-group target. However, if such
evaluation is purely based on an analysis of the network structure,
the resulting distribution of phases must be disregarded. This has
the drawback that we do not know {\it a priori} which oscillators
belong to each group.

To tackle this problem we have adapted two algorithms of community
detection in networks \cite{Field,Newm}, whose aim is to identify
groups of network nodes which are internally best connected. In our
case, this task is translated into the identification of two groups
which maximize the number of attractive interactions inside each
group. The result is given as an ordering of the oscillators, with
lower ranks corresponding to one group and higher ranks to the
other. In the first method, called spectral partitioning algorithm
\cite{Field}, the size of the two groups is fixed --in our case, to
$N_1=15$ and $N_2=35$. In the second, which is based on the
maximization of the network modularity \cite{Newm}, the group size
is free.

\begin{figure}[h]
\begin{center}
\resizebox{\columnwidth}{!}{\includegraphics{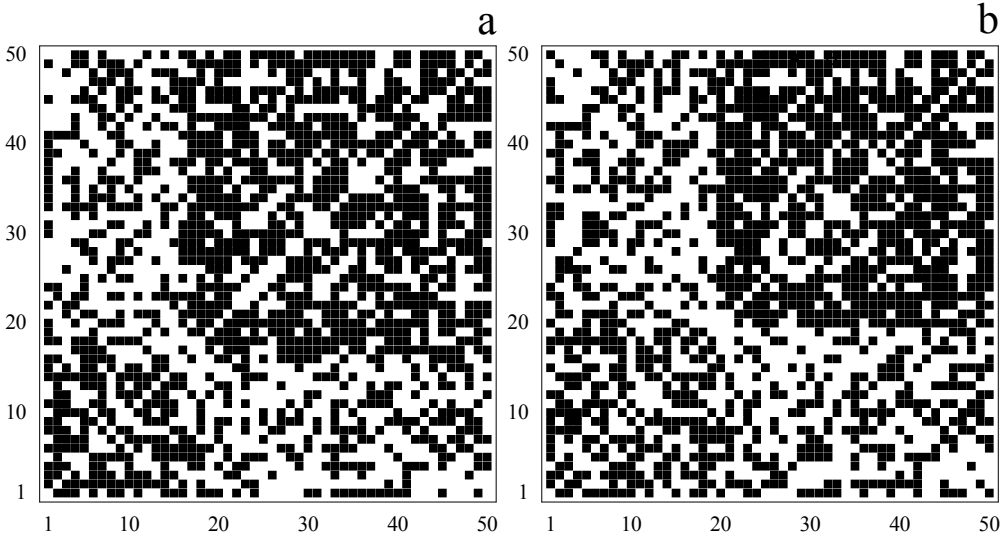}}
\end{center}
\caption{Graphical representation of the adjacency matrices of the
network of attractive interactions, after detection of the two
groups, (a) by the algorithm of spectral partitioning  and  (b) by
maximizing the network modularity.} \label{f4x}
\end{figure}

Figure \ref{f4x}a is a graphical representation of the adjacency
matrix of the network of attractive interactions with the numbering
resulting from the first method. Squares represent oscillator pairs
with attractive interactions. In the ideal case where both the
adaptation algorithm and the ordering method are perfectly
successful, all squares should be concentrated in two
non-overlapping blocks of sizes $N_1\times N_1$ and $N_2\times N_2$,
situated along the diagonal. Though our result is far from this
ideal configuration, a larger density of squares is apparent over
the two blocks. Figure \ref{f4x}b presents the result of using the
second method, with free group sizes. The concentration in the
expected zones is again clear, though the resulting groups are of
slightly different sizes: $N_1 \approx 20$ and $N_2\approx 30$.

In order to quantify the performance of adaptation at the level of
the network structure, we may use the results depicted in Fig.
\ref{f4x} to introduce an index which measures to what extent has
the adjacency matrix achieved the ideal division into two blocks.
Let $f$ be the fraction of sites inside the blocks (excluding the
diagonals) which are occupied by squares after adaptation, and
$f_R=1-x$ the fraction of occupied sites for a random distribution
of attractive interactions. Note that $f_R$ coincides, up to random
fluctuations, with the fraction of occupied sites inside the blocks
in the initial condition. We define
\begin{equation}
D = \frac{f-f_R}{1-f_R} .
\end{equation}
In the ideal final state, we would have $D=1$, because the filling
of the two blocks would be perfect ($f=1$). Meanwhile, for randomly
distributed interactions, $f\approx f_R$ and $D\approx 0$. Taking
into account the effect of fluctuations, in fact, we find $D\sim
N^{-1}$. The values of $D$ calculated for the results shown in Fig.
\ref{f4x} are (a) $D=0.26$  and (b) $0.33$. As for the case of the
index $C$, Eq. (\ref{C}), these values are considerably smaller than
the ideal. However, during adaptation, they have grown by an order
of magnitude, which represents an important improvement towards the
target interaction pattern.

\section{Discussion and conclusion}

For the $50$-oscillator ensemble, we have repeated the analysis with
groups of various sizes ($N_1=10$, $20$, $25$) and a different
coupling strength ($K=8$), and have always obtained consistent
results --not presented here for conciseness. The degree of
adaptation both in dynamics and in network structure, as measured by
the quantities $C$ and $D$, was similar to that presented above. We
have also tested the adaptation algorithm using the difference
$\zeta=z_2^2-z^2$, instead of the two-cluster order parameter $z_2$,
as the macroscopic measure of collective organization. In view of
the results displayed in Fig. \ref{f3}, this criterion is expected
to exhibit better performance for large values of $K$, where the
difference between the two order parameters $z$ and $z_2$ is large
for the two-group configuration, and small for the random
interaction network. Finally, we have verified that adding a
moderate number of oscillators with random interactions, once the
original ensemble has already undergone the adaptation process, does
not produce a substantial change in the collective dynamical state
of the system. All these tests point at the robustness of the
results presented in Section \ref{adapta}.

Regarding the collective dynamical state achieved by the ensemble
after the adaptation process, it is interesting to emphasize that,
at the moderate coupling strength considered here, a difference
results between the target state of two clusters with opposite
phases, on one hand, and ensemble synchronization, on the other. As
a result of adaptation, in fact, most oscillators end in pairs of
clusters with opposite phases --indicating that an important
fraction of repulsive interactions has been redistributed as
expected. However, different pairs of clusters have different
average frequencies and, therefore, are not mutually synchronized.
In other words, there is a certain degree of independence between
synchronized clusters and opposite-phase clusters, as depicted in
Fig. \ref{f5}. This ``multilevel'' clustering may represent a rich
regime from the viewpoint of the collective dynamics \cite{pnas},
and deserves further study.

As for the network structure, the results shown in Fig. \ref{f4x},
along with the values obtained for the quantity $D$, may seem a
modest achievement for a Monte Carlo adaptation algorithm on a
$50$-oscillator ensemble. One should take into account, however,
that the algorithm explores the space of possible distributions of
$N_1 N_2$ repulsive interactions over the $N(N-1)/2$ oscillator
pairs. For $N_1=15$ and $N_2=35$, identifying the ideal two-group
configuration amounts at singling a particular state out of $4.7
\times 10^{361}$ possibilities! This huge number approximately
equals $4^{600}$, which indicates that the task is comparable to
individualizing a  specific $600$-nucleotide long genetic message.
Note that this figure is not far below the length of some short
coding sequences in real genomes \cite{geno}.

We have shown in this paper that the distinctive scale separation
between the mechanisms of mutation and selection which characterize
biological evolution  --or, equivalently, neural and supervising
mechanisms in learning processes-- can be emulated by an ensemble of
very simple interacting dynamical systems, which adapts to a
prescribed form of collective behaviour by gradual changes in its
interaction pattern. Admittedly, the two-cluster synchronization
state here chosen as the adaptation target is by no means as complex
as some of the most elementary functional configurations of
biological entities. However, as long as a quantification of the
adaptation level  --analogous to our two-cluster order parameter
$z_2$-- is identified, the same algorithm could be in principle
implemented for evolution towards more sophisticated collective
dynamics. The same consideration can be applied to systems whose
elementary components are dynamically richer that the phase
oscillators studied here, such in the case of chaotic units
\cite{zhou}, or whose interaction pattern is structurally closer to
those observed in real biological networks --typically, much more
sparse than our fully connected graphs. Additional dynamical
ingredients may include, for instance, the effects of internal
fluctuations and noise.

In any case, the quantitative extent and limitations of the
correspondence between biological evolution and adaptation in
networks of dynamical elements --in particular, with respect to the
performance of the adaptation algorithm-- as well as the possibility
of implementation in other complex systems with different forms of
self-organizing collective dynamics, is worth considering in future
work.

\end{document}